\documentclass[cits]{PoS}

\title{Models for Microquasars}

\ShortTitle{Models for Microquasars}

\author{\speaker{Julien Malzac}\\
        CESR, OMP, UPS, CNRS;  Toulouse, France\\
        E-mail: \email{malzac@cesr.fr}}

\abstract{
I review current models used to interpret the spectra and variability of microquasars. 
Among other things, I discuss the structure of the accretion flow and its dependence on mass accretion rate,  the intrinsic connection between hot comptonizing corona and compact radio jet in the hard state, as well as  possible models for the spectral hysteresis observed during outbursts of transient sources.  Finally I comment on several models for the non-poissonian X-ray noise in black hole binaries which, at least in some instances, is suspected to be associated with some form of coupling between disc and jet activity.  
}

\FullConference{VI Microquasar Workshop: Microquasars and Beyond\\
		 September 18-22, 2006\\
		 Como, Italy}

\begin{document}

\section{X-ray spectral states and the structure of the accretion flow}\label{sec:xray+geo}

\begin{figure}[t]

\includegraphics[width=0.55\textwidth]{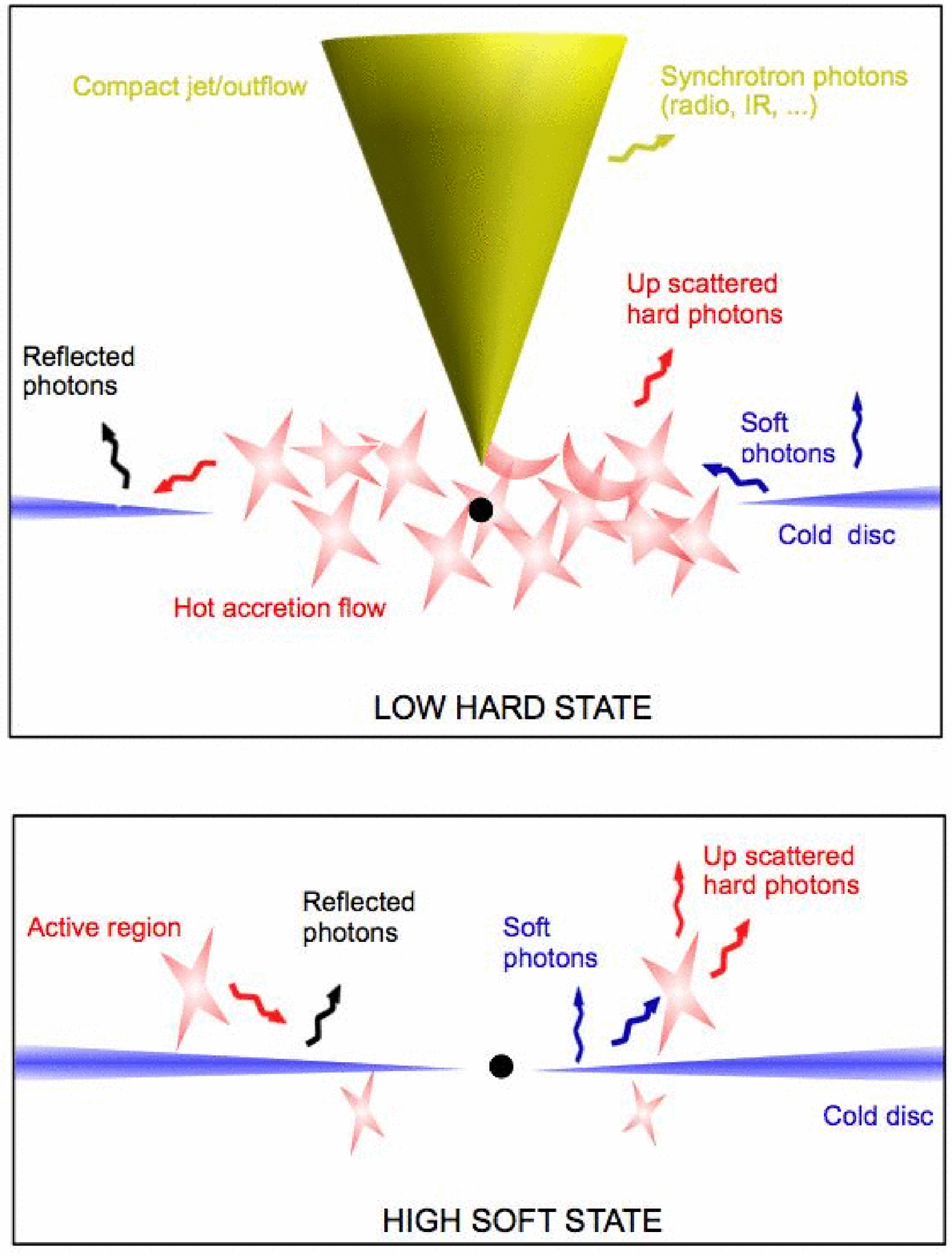}
\resizebox{0.5\textwidth}{!}{\parbox{9cm}{\vspace{-12cm}\hspace{+0cm}\includegraphics[width=8cm]{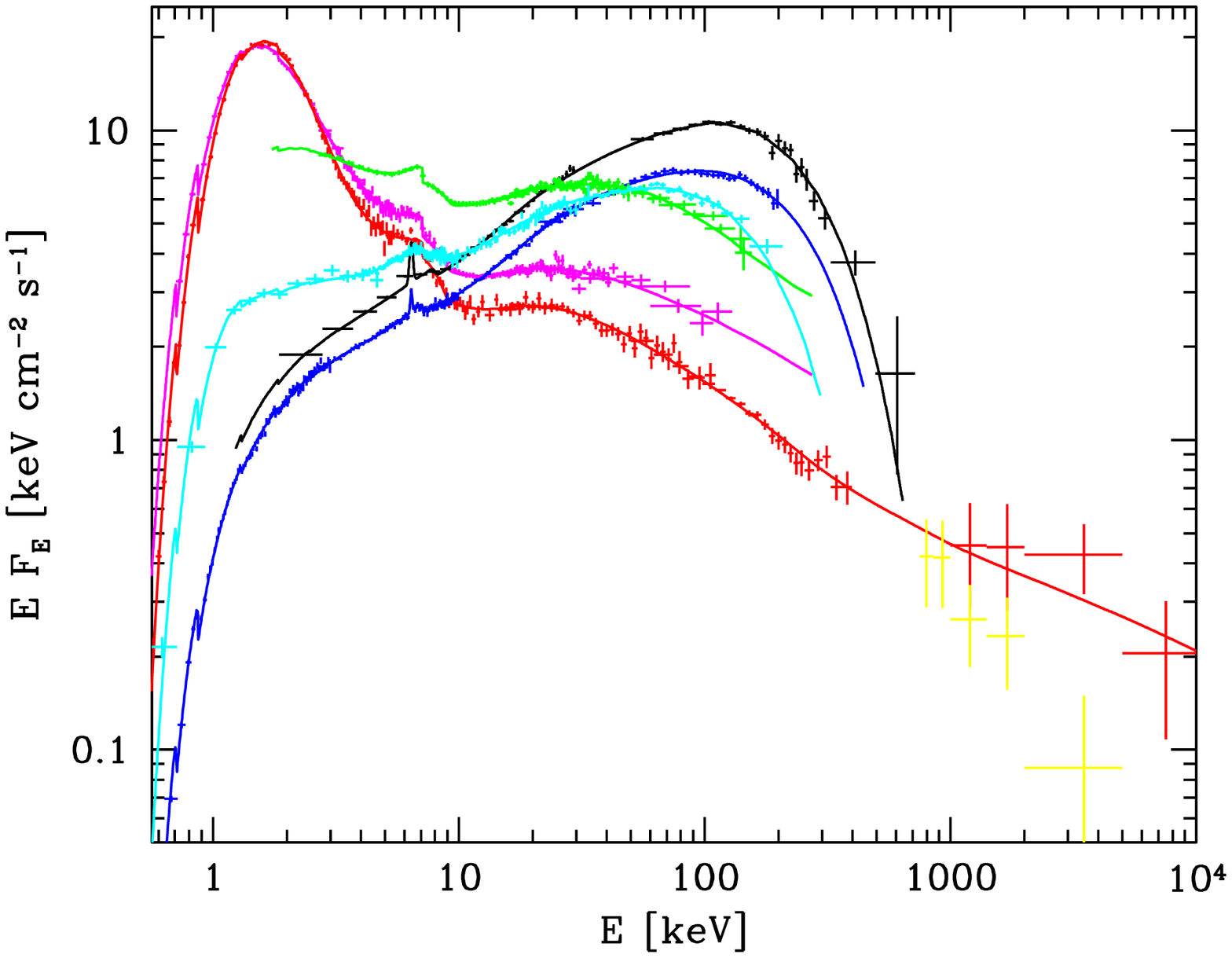}\\ \hspace*{0cm}\includegraphics[width=8cm]{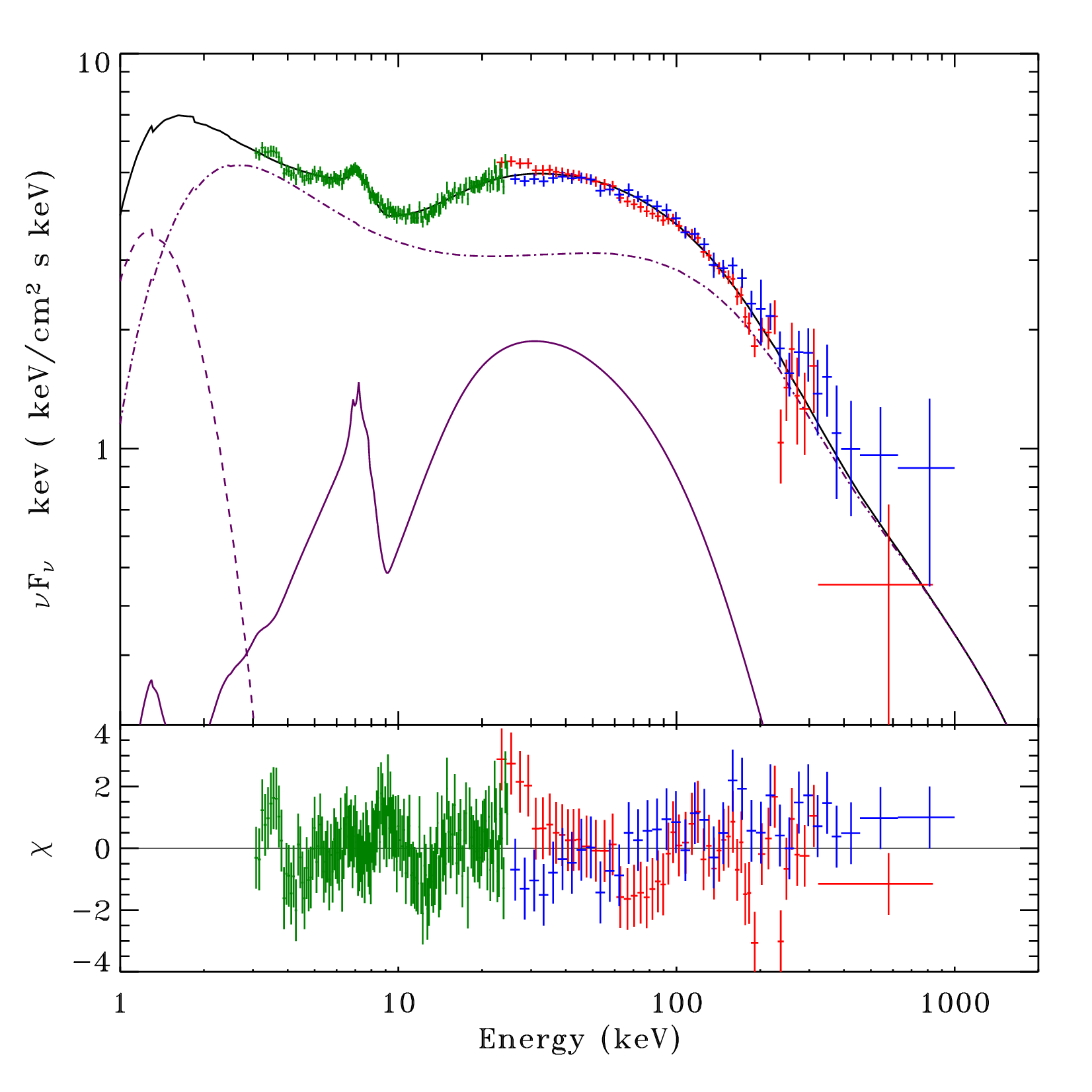}}}
\caption{Left hand side panels: Structure of the accretion flow during LHS (top) and HSS (bottom) according to the standard scenario. Right hand side panels: Observed spectra of Cygnus X-1 . Upper panel (from \cite{Zdziarski et al 2002}):  the LHS (black and  blue), HSS (red, magenta), IMS (green, cyan). The solid curves give the best-fit Comptonization models (thermal in the hard state, and hybrid, thermal-nonthermal, in the other states). Lower panel:  Time averaged INTEGRAL spectrum of Cygnus X-1 
during a mini-state transition (intermediate state).
 The data are fitted with the 
thermal/non-thermal hybrid Comptonisation model {\sc eqpair}
 with \emph{mono-energetic} injection of relativistic electrons. 
 The lighter curves show the reflection component (solid), 
 the disc thermal emission (dashed) and the Comptonised emission (dot-dashed).
 The green, red and blue crosses show the {\it JEM-X},
  {IBIS/ISGRI} and {SPI} data respectively. See \cite{M06} for details.
\label{fig:cygx1spectra} }
\end{figure}

Most of the luminosity of accreting black holes is emitted in the X-ray band. This X-ray emission is strongly variable.  A same source  can be  observed with very  different X-ray spectra (see e.g. 
 \cite{zg04}).  Fig.~\ref{fig:cygx1spectra} shows various spectra from the prototypical source  Cygnus~X-1 observed at different epochs. There are two main spectral states that are fairly steady and most frequently observed.
At high luminosities the accretion flow is in the High Soft State (HSS), caracterized by a strong thermal disc and reflection components (believed to be due to X-ray illumination of the accretion disc) and a weak and non-thermal component believed to be produced through  non-thermal  (or hybrid thermal/non-thermal) comptonisation in a hot corona. At luminosities lower than a few percent of Eddington ($L_{Edd}$), the sources  are generally found in the Low Hard State (LHS) in which the disc blackbody and reflection features are much weaker, while  the corona has a thermal distribution of comptonising electrons and dominates the luminosity output of the system. Beside the LHS and HSS, there are several other spectral states that often appear, but not always, when the source is about to switch from one of the two main states to the other. Those states are more complex and difficult to define. We refer the reader to \cite{McClintock and Remillard 2006} and \cite{Belloni et al 2005} for two different  spectral classifications based on X-ray temporal as well as spectral criteria and radio emission \cite{Fender 2006}.  In general, their spectral properties  are intermediate between those of the  LHS and HSS.

The different spectral states are usually understood in terms of changes in the geometry of the accretion flow.  In the HSS, according to the standard scenario sketched in Fig.~\ref{fig:cygx1spectra},  a standard geometrically thin disc extends down to the last stable orbit and is responsible for the dominant thermal emission. The non-thermal hard X-ray emission is generally believed to originate from an Accretion Disc Corona (ADC) constituted of small active regions located above and below the disc. Due to magnetic buoyancy, the magnetic field lines rise above the accretion disc, transporting a significant fraction of the accretion power into the corona where it is then dissipated through magnetic reconnection \cite{Galeev et al 1979}.  This leads to particle acceleration in the ADC.  A population of  high energy electrons is formed which then cools down by up-scattering the soft photons emitted by the disc. This produces the high energy non-thermal emission which in turn illuminates the disc forming strong reflection features (see e.g. \cite{zg04}). 
There are no strong alternatives to ADC model for the soft state. 
The most relevant is probably the bulk motion model  \cite{Laurent and Titarchuk 2001} which attributes  the hard X-ray emission  to Comptonisation in the innermost  region by the plasma inflowing into the black hole.
However, as pointed out in \cite{Nied{\'z}wiecki and Zdziarski 2006}, this model  predicts  a sharp cut-off in the high energy spectrum around 100 keV,  and therefore does not explain the MeV emission observed in several HSS sources.
In the LHS, the standard geometrically thin disc does not extend to the last stable orbit, instead, the weakness of the thermal features suggests that it is truncated at distances ranging from a few hundreds to a few thousands gravitational radii from the black hole (typically  1000--10000 km). In the inner parts, the accretion flow takes the form of a hot geometrically thick, optically thin disc. A solution of such hot accretion flows was first proposed by Shapiro et al \cite{SLE76}. In these hot acccretion flows the gravitational energy is converted in the process of viscous dissipation into the thermal energy of ions. The main coupling between the electrons and the ions is Coulomb collision which is rather weak in the hot thin plasma. Since radiative cooling of the ions is much longer than that of the electrons, the ions temperature is much higher than the electron temperature.  This two temperature plasma solution  is  thermally unstable \cite{Pringle 1976}  but can be stabilized if advection of the hot gas into the black hole  dominates the energy transfert for ions as in advection dominated accretion flows (ADAF) solutions  (see \cite{Ichimaru 1977}, \cite{Narayan and Yi 1994}, \cite{Abramowicz et al 1996}). 
Emphazising the fact that in the ADAF solution the accreting gas has a positive Bernouilli parameter, Blandford and Begelman \cite {Blandford and Begelman 1999} proposed a variant of the ADAF model in which a significant fraction of the accreting material is advected into an outflow rather than in the black hole (ADIOS). In these hot accretion flow solutions most of the accretion power is either swallowed by the black hole or converted into kinetic power of an outflow. They are therefore radiatively inefficient accretion flows (RIAF).  Yuan \cite{Yuan 2001}  has shown that a hot flow may also exist  in the higher accretion rate regime where the coupling between electrons and ions becomes effective: the flow is then radiatively efficient. 
In all these hot flow models the electrons have a thermal distribution and cool down by Comptonisation of the soft photons coming from the external geometrically thin disc, as well as IR-optical photons internally generated through self-absorbed synchrotron radiation.  The balance between heating and cooling determines the electron temperature which is found to be of the order of 10$^9$ K, as required to fit the spectra.
The weak reflection features of the LHS are produced through illumination of the cold outer disc by the central source.  
By extension, the hot accretion flow of the LHS  is frequently refered to as  "corona" despite the lack of a direct physical analogy with the  rarefied gaseous envelope of the sun and other stars or the ADC of the HSS. 

\begin{figure}[tb]
%\centerline{\psfig{file=sed_+mod.ps,angle=90.,width=9.cm}}
\resizebox{0.6\textwidth}{7cm}{\rotatebox{90}{\includegraphics{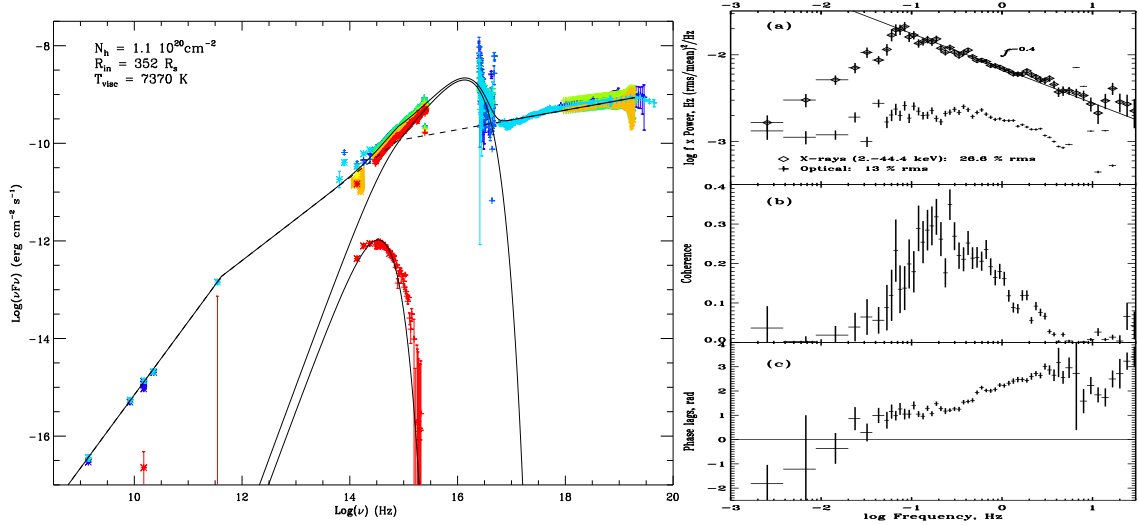}}}  \resizebox{0.4\textwidth}{7cm}{\includegraphics{3498.f5}}
\caption{\label{sed_tout+mod}  Left: Spectral Energy Distribution of XTE J1118+480 during its 2000 outburst (from \cite{C03}). Right: The optical/X-ray correlations  of XTE J1118+480  in the Fourier domain.
 {\bf a)} X-ray and optical power spectra. The counting noise was
subtracted.
 {\bf b)} X-ray/optical
coherence. {\bf c)} phase-lags as function of
Fourier frequency. A positive lag implies 
that the optical is delayed with respect to the X-rays (from \cite{M03}).
\label{fig:sedxte}}
\end{figure}

\begin{figure*}[t]
\center
\resizebox{0.6\textwidth}{!}{\includegraphics[clip=true]{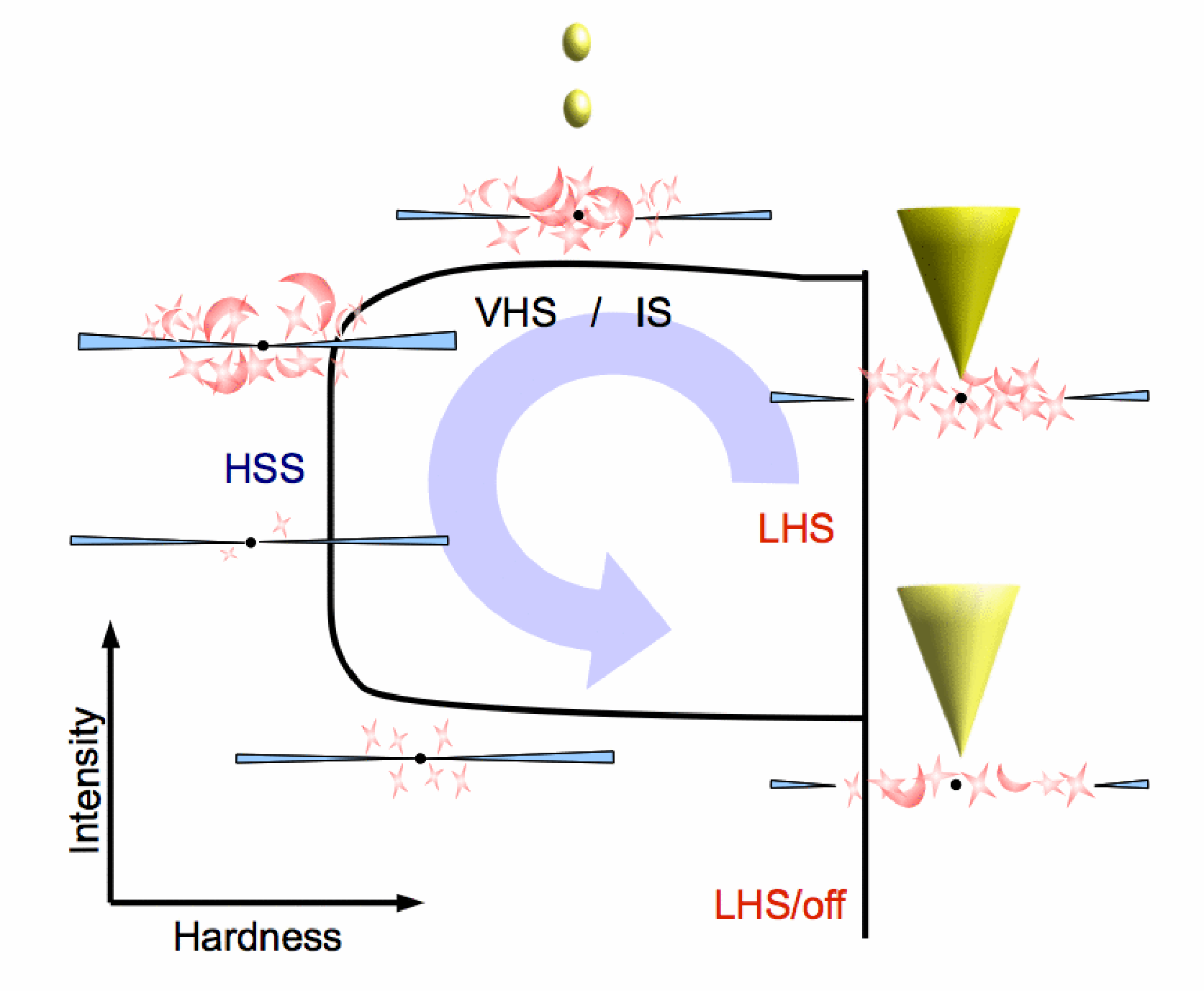}}
\caption{
\footnotesize
The black curve shows the typical path followed by  a black hole X-ray transient in the hardness vs intensity diagram during an outburst. The various sketches illustrate the standard scenario for the evolution of the geometry of the corona-disc-jet system along this path (see text).}
\label{fig:hyste}
\end{figure*}

\section{An alternative model for the LHS: accretion disc corona}

The presence of the hot accretion disc is not the only possibility to explain the LHS. 
 It was suggested long ago that instead, a real ADC system similar to that of the HSS but with a pure Maxwellian distribution of comptonising electrons, may reproduce the spectra as well \cite{Bisnovatyi-Kogan and Blinnikov 1976} \cite{Liang and Price 1977}. This would imply a cold geometrically thin disc extending down very close to the black hole in the LHS. Unlike in the HSS, this disc does not produce a strong thermal component in the X-ray spectrum because it is too cold. Indeed, it has a much lower temperature because most of the accretion power is not dissipated in the disc, instead, it is assumed to be transported away to power a strong ADC and the compact jet. This, at least in principle, could be achieved either through transport via buyonancy of the magnetic field \cite{Miller and Stone 2000}\cite{Merloni and Fabian 2002} or through the torque exerted on the accretion disc by a strong magnetic field threading the disc and driving the jet \cite{Ferreira  1997}. 

Due to the complexity of the ADC physics it is not possible to obtain simple analytical solutions predicting the main properties of the ADC unless  some parametrization of the energy transfert between the disc and the corona is used (see e.g.  \cite{Svensson and Zdziarski 1994}).
However, in practice, the appearance of the ADC does not depend on the details of the energy transport and dissipation mechanisms. Indeed, Haard and Maraschi \cite{Haardt and Maraschi 1991} pointed out the existence of a strong radiative feedback between the cold disc and the hot corona. A fixed fraction of the power dissipated in the corona in the form of hard X-ray photons intercepts the cold disc where it is absorbed. As a consequence the disc heats up, and the absorbed energy is finally reemitted in the form of thermal soft photons. A fraction of those soft photons re-enters the corona providing the major cooling effect to the corona trough inverse Compton. Due to this strong feedback from the disc, the cooling rate of the corona scales like the heating rate. The coronal temperature is determined only by the geometry of the ADC (that controls the fraction of coronal power that returns to the corona in the form of soft photons). For instance, an extended ADC sandwiching the cold accretion disc, would intercept more cooling  photons from the accretion disc than a patchy corona made of a few compact  active regions covering a small fraction of the disc, and therefore would have a lower temperature and a softer comptonised X-ray spectrum. 
During the nineties several groups  \cite{Haardt and Maraschi 1993} \cite{Stern et al 1995a}\cite{Poutanen and Svensson 1996}
 performed detailed computations of the resulting equilibrium spectra for various geometries. 
They concluded that the ADC has to be patchy in order to produce spectra that are hard enough \cite{Haardt et al 1994}\cite{Stern et al 1995b} even when the effects of ionisation on the disc albedo are accounted for \cite{Malzac et al 2005}.
Yet, those ADC models had an important problem:  the production of an unobserved  strong thermal component due to reprocessing of the radiation illuminating the disc. Indeed, in the case of an active region emitting isotropically, about half of the luminosity intercepts the disc and the thermal reprocessing component is comparable in luminosity to the primary emission.  Actually, in observed LHS spectra this thermal disc component is so weak that it is barely detectable. Morevover, the amplitude of the reflection features observed in the LHS are lower than what is expected from an istotropic corona by at least a factor of 3, in some cases they are so weak that they are not even detected. 
This led to consider models where the coronal emission is not isotropic. It was suggested that the corona is unlikely to stay at rest with respect to the accretion disc  \cite{Beloborodov 1999}. Due to the anisotropy of the dissipation process or simply due to radiation pressure from the disc, the hot plasma is likely to be moving at mildly relativistic velocities. Then, due to Doppler effects, the X-ray emission is strongly beamed in the direction of the plasma velocity. In the case of a velocity directed away from the accretion disc (outflowing corona), the reprocessing features (both reflection and thermalised radiation)  are strongly suppressed. Moreover, due to the reduced feedback from the disc, the corona is hotter, and harder spectra can be produced. 
Detailled non-linear Monte-Carlo simulations of this dynamic ADC equilibrium are presented in \cite{Malzac et al 2001} . A comparison of these results with observations showed that compact active regions of aspect ratio of order unity,  outflowing with a velocity of 30 percent of the speed of light could reproduce the LHS spectrum of Cygnus X-1. 
Moreover, since the velocity of the coronal plasma controls both the strength of the reflection features and the feedback of soft cooling photons from the disc, it predicts a correlation between the slope of the hard X-ray spectrum and  the amplitude of the reflection component. 
Such a correlation is indeed observed in several sources  \cite{Zdziarski et al 2003} and is well matched by this model.
Recently the ADC models for the LHS obtained some observationnal  support, with the discovery of a relativistically broadened iron line in the LHS of GX339-4 \cite{Miller et al  2006}. Such relativistically broadened lines require disc illumination taking place very close to the black hole. This observation suggests that, at least in some cases, a thin disc is present at, or close to, the last stable orbit in the LHS.
Incidentally, in the framework of the outflowing ADC model, the velocity of the corona required to fit hard state spectra appears to be comparable to the estimates of the compact radio jet velocity \cite{Gallo et al 2003}. This suggests a direct connection between  corona and jet.

\section{The jet corona connection}\label{sec:corjetcoup}

Multi-wavelength observations
of accreting  black holes in the LHS have shown the presence of
an ubiquitous flat-spectrum radio emission (see e.g \cite{Fender 2006}), that may
extend up to infrared and optical wavelengths (see Fig.~\ref{fig:sedxte}). 
The properties of the radio emission indicate it is likely  produced
by synchrotron emission from relativistic electrons in compact,
self-absorbed jets  \cite{Blandford and Konigl 1979} \cite{Hjellming and Johnston 1988}. 
This idea was confirmed by the discovery 
of a continuous and steady milliarcsecond 
compact jet around Cygnus X-1 \cite{Stirling et al 2001}.
Moreover, in LHS sources a tight 
correlation has been found between the hard X-ray and radio luminosities, holding over more than three decades in luminosity 
 \cite{Gallo et al 2003}.
In contrast, during HSS episodes the sources appear to be
radio weak \cite{Corbel et al  2000}, 
suggesting that the Comptonising medium of the low/hard 
state is closely linked to the continuous ejection of matter in 
the form of a small scale jet. 
When the importance of the connection between radio and X-ray emission was realised,
it was proposed that the hard X-ray emission could be in fact synchrotron emission in the jet, rather  than  comptonisation in a hot accretion flow/corona \cite{Markoff et al 2001}. However, it seems that in most sources the synchrotron emission alone is not enough to reproduce the details of the X-ray spectra. In the most recent version of this model a thermal Comptonisation component was added which appears to provide a dominant contribution to the hard X-ray spectrum \cite{Markoff et al 2005}.  This component is supposedly located at the base of the jet which forms a hot plasma very similar to an ADC.  

In the context of ADC/hot disc models the correlation between X-ray and radio emissions, simply tells us that the corona and the compact jet of the LHS are intimately connected. 
A strong corona may be necessary to launch  a jet and/or could be  the physical location where the jet is accelerated or launched \cite{Merloni and Fabian 2002}.  
Jet spectral components have  recently been added to the hot disc models and make it possible to produce good fits of the whole spectral energy distribution from radio to hard X-rays \cite{Yuan et al 2007}, although the physics of the jet/ADAF connection has not been worked out yet.
Following a different approach, K\"{o}rding et al. \cite{Koerding et al 2006} show that the observed X-radio correlation can be reproduced provided that the hot flow/corona is radiatively inefficient (i.e. luminosity scales like the square of mass accretion rate) and that a constant fraction of the accretion power goes into the jet (i.e. jet power scales like mass accretion rate). 
However there is presently no detailed model to explain how the physical connection between jet and corona operates.  A possible basic explanation was proposed by Meier \cite{Meier 2001} for ADAF like accretion flows and later extended to the case of ADC by Merloni and Fabian \cite{Merloni and Fabian 2002}.  It goes as follows. Models and simulations of jet production  indicate that jets are driven 
by the polo{i}dal component of the magnetic field \cite{Blandford and Znajek 1977}\cite{Blandford and Payne 1982}\cite{Ferreira 1997}. If we assume that the magnetic field is generated by dynamo processes in the disc/corona, the strengh of the poloidal component is limited by the scale height of the flow \cite{Livio et al 1999}\cite{Meier 2001}\cite{Merloni and Fabian 2002}. 
Therefore geometrically thick accretion flows should be naturally more efficient at launching jets.

Despite this strong link between the corona and outflow in the LHS,
there are indications that in Intermediate States (IS) the jet  is connected to the accretion disc rather than the corona. For instance, Malzac et al. \cite{M06} report the results of an observation of Cygnus X-1 during 
a mini-state transition (see Fig.~\ref{fig:cygx1spectra}). The data indicate that the radio jet luminosity is anti-correlated with the disc luminosity and unrelated to the coronal power. This is in sharp contrast with previous results
obtained for the LHS, and suggests a different mode of coupling between the jet, the cold disc, and the corona in Intermediate States. The reason for the anti-correlation is most probably  that  state transitions are associated with a redistribution of the available accretion power between the compact jet and the cold accretion disc. In the standard scenario described in section~\ref{sec:xray+geo}, this redistribution of accretion power could occur because the jet  shrinks as the inner radius of the outer disc moves closer to the black hole.    

\section{Evolution of the geometry during outburst: the hysteresis problem}

Transient black hole binaries are a class of X-ray binaries that are detected  only occasionally when  they show some period of intense activity (outbursts) lasting for a few months (see e.g. \cite{Tanaka and Shibazaki 1996}). These outbursts have  recurrence times ranging from years to dozens of years. Between two episodes of activity the source is in an extremely faint quiescent state. During an outburst, the source luminosity  varies by many orders of magnitude. The study of the spectral evolution during outbursts is perfectly suited to study the dependence of the structure of the accretion flow  on the mass accretion rate. This spectral evolution is conveniently described using hardness intensity diagrams. Those diagrams show the evolution of the source luminosity as a function of spectral hardness (defined as the ratio of the observed count rates in a high to a lower energy band).
 Typically, during outbursts the sources follow a \emph{q}-shaped path in the hardness intensity diagram (see Fig.~\ref{fig:hyste}). Initially in quiescence with a very low luminosity, they are in the LHS with a high hardness.  In the standard scenario\footnote{We note that a different geometrical interpretation of the outburts cycle in terms of a jet dominated accretion flow was recently proposed by Ferreira et al. \cite{Ferreira et al 2006}.
}, the cold disc is truncated very far away from the black hole, and the emission is dominated by the hot accretion flow. Then, as luminosity increases by several orders of magnitude the hardness remains constant.  
 The structure of the accretion flow is stable, the  cold disc inner radius probably moving slowly inward.  During this rise in X-ray luminosity  a compact jet is present and its radio emission  correlates with the X-ray emission. At a luminosity that is always above a few percent of Eddington,  the inner radius of the cold accretion disk decreases quickly.
 This reduction of the inner disc radius is associated with either
  the cold disk penetrating inside the hot inner flow, or the  later collapsing into an optically thick accretion disk  with small active regions of hot plasma on top of it 
   (\cite{Zdziarski et al 2002}). In both cases the enhanced soft photon flux from the disk 
   tends to cool down the hot phase, leading to softer spectra. The source therefore moves from the right toward  the left hand side at almost constant luminosity.  During this hard to soft  state transition  the sources usually show a strong coronal activity (both the disc and corona are strong), this is also during this transition that optically thin radio flares associated with relativistic and sometimes superluminal ejections are observed. This region of the hardness intensity diagram is often called the Very High State (VHS). 
 Once a source has reached the upper left hand corner of the hardness intensity diagram, 
 it  never turns back to the LHS following the same path. Instead when the luminosity decreases, it stays in the HSS and goes down vertically and  the strength of the non-thermal corona decreases as well.
 This goes on until the luminosity reaches $\simeq$ 0.02 percent of Eddington. Then it moves horizontally  to the right,  back to the LHS, and then down vertically to quiescence.

It is to be noted that the luminosity of state transition from soft to hard seems to be relatively fixed around 
0.02 percent of Eddington \cite{Maccarone 2003}. On the other hand the hard to soft transition can vary, even for the same source, by several orders of magnitude and it is always higher than the soft to hard transition luminosity. 
The question of what controls the different transitions luminosity is still open. Indeed, such a complex behaviour is not expected a priori. In all accretion models, there is only one relevant externally imposed parameter: the mass acretion rate. It was not anticipated that for the same 
luminosity the accretion flow could be in different states. It was recently suggested that this complex behaviour could be explained by the existence of a second parameter that could be the magnetic flux advected with the accreting material \cite{Spruit and Uzdensky 2005}.
But the most detailed explanations so far are based on the idea that this second parameter is nothing else than the history of the system. In other word the accretion flow would present an hysteresis
behaviour because going from LHS to HSS is not the same thing as going from HSS to LHS. Within this line of reasoning, a promising idea  was proposed by Meyer-Hofmeister et al \cite{Meyer-Hofmeister et al 2005}
 who suggested that the hysteresis could be linked to a condensation/evaporation equilibrium in an ADC system. They assume an advective corona on top of a thin optically thick accretion disc and  allowing for mass exchange between the corona and the disc. Depending on the temperature and density in the corona and the disc, the material in the corona may condensate into the disc and the disc may evaporate. Meyer-Hofmeister et al \cite{Meyer-Hofmeister et al 2005}  have computed the resulting equilibrium between an accretion disc and a corona as a function of the distance to the black hole and mass accretion rate (see also  \cite{R{\'o}{\.z}a{\'n}ska and Czerny 2000} \cite{Mayer and Pringle 2007}).
 They have found that at large distance the conditions are such that an equilibrium solution can always be found. At  closer distances from the black hole, disc evaporation becomes more and more important and, under some circumstances there is a critical radius below which the disc fully evaporate and no thin disc is possible. Below this radius we are then left with a pure advection dominated hot accretion flow. The location of this transisition radius depends on the mass accretion rate. At low mass accretion rate the transition radius is at 10$^3$ to 10$^4$   Schwarschild radii and decreases slowly as the accretion rate increases until the critical mass accretion of about a few percent of Eddington is reached. Then, the inner radius suddenly drops to the last stable orbit: this is the hard to soft  state transition. As long as the source is in the LHS the flux of soft cooling photons in the corona is weak, and,  as a consequence the corona is hot and evaporation dominates over condensation. Once the HSS is reached however, there is a disc down to the last stable orbit emitting a strong thermal soft emission that efficiently cools the corona, keeping the condensation process strong even when the mass accretion rate decreases. As a consequence the transition from the soft to LHS occurs at a luminosity that is  about one order of magnitude lower  than the hard to soft transition. 

Although this model so far provides the only quantitative explanation of the hysteresis behaviour,  it is not perfect.  First, the presence of the jet, possibly impacting on the disc/corona equilibrium is completely ignored. Second, this model does not explain why the transition LHS to HSS can occur at different luminosities (the transition luminosity  is quite constrained), nor the absence of hysteresis behaviour in persistent sources such as Cygnus X-1. A second independent parameter may still be required.

\section{Timing properties}

X-ray binaries harbour a strong rapid X-ray variability which presents a very complex and richly documented phenomenology \cite{van der Klis 2006}. Modeling this fast variability is very challenging. The main problem that any model must overcome is a time-scale problem. Indeed, most of the high energy photons (and variability) must originate deep in the potential. The variability is therefore expected to occur on time-scales of the order of the dynamical time-scale close to the black hole, which is of order of 1 ms for a ten solar mass black hole. On the other hand, observations show that most of the X-ray variability occurs on time-scales ranging from 0.1 to 10 s (see e.g. panel \emph{a} on the right hand side of Fig.~\ref{fig:sedxte}). In comparison there is virtually no variability on time-scales shorter than 0.01 s.  Thinking in terms of Power Density Spectrum (PDS), the observed power requires short distances from the black hole, but the observed frequencies imply large distances. 

Another intriguing feature of the rapid variability of black hole binaries is the existence of delays between energy bands (the hard photons lagging behind the softer photons \cite{Miyamoto and Kitamoto 1989}). These time-lags are often believed to be associated to the time required for a photon to gain energy through multiple compton scaterring as it travels in the corona \cite{Kazanas et al 1997}.  But this explanation is probably wrong:
First,  the observed time-lags are very long and would require an unplausibly large Comptonizing corona ($\sim$ 10000 Rg). But more fundamentally, the time-lags cannot be due to Comptonisation delays because this requires the variability to be driven by fluctuations of the soft seed photon luminosity  while the hot  corona keeps constant physical parameters.  Indeed, Malzac and Jourdain  \cite{Malzac and Jourdain 2000} have shown that in a real situation the corona responds to the fluctuations of the soft photon field and damps out any variability. For example, if the soft photon density is suddenly increased while the dissipation rate in the corona is kept constant, then the temperature drops almost instantaneously (on a Compton  cooling time scale which is much shorter than the light crossing time). In a fluctuating soft photon field, the coronal temperature tends to adjust very quickly to keep the luminosity constant. Models in which the variability is driven by the soft photon field might be successful in reproducing the shape of the power spectra or time-lags (see e.g. \cite{B{\"o}ttcher 2001}) but they \emph{cannot} produce the strong variability rms amplitudes of 10 to 40 percent that are typically observed. 
Therefore the rapid variability must be driven by  changes in the physical parameters of the corona (most likely the hot plasma heating rate), rather than by fluctuations of the soft seed photon input.

There are several models that successfully reproduce the PSD shapes and amplitude as well as other observed properties such as the time-lags. So-called 'shot noise models'  (see e.g. \cite{Lochner et al 1991}) attempt to phenomenologically describe the light curves as a succession of random flares with specific time scales, amplitudes, and occurence rates. More physically motivated models may involve turbulence in accretion discs \cite{Nowak and Wagoner 1995}, avalanche of magnetic flares \cite{Poutanen and Fabian 1999}, or MHD in the plunging region \cite{Hawley and Krolik 2001}. 
Sometimes, the time scale problem is solved by invoking a mechanism correlating short time-scale fluctuations to generate the observed long time-scale fluctuations. However it was shown by \cite{Uttley et al 2005} that the variability is non-linear. 
Indeed, they demonstrated the existence of a linear correlation between the measured count rate and the rms variability that seems to be present in all sources and in all spectral states and on all time-scales. This correlation shows that the short time-scales are modulated by the longer time scales. This appears to favour models where the variability is built from large to short times-scales.

 A possibility could be that fluctuations are generated at large distance 
and propagate inward. For instance, Lyubarskii \cite{Lyubarskii 1997} postulates fluctuations in viscosity on local inflow timescale (itself viscous) over a wide range of distances $r$ from the black hole
and computes the resulting fluctuations of the mass accretion rate at the inner disc radius.  For fluctuations independent of $r$, this gives a power spectrum $p( f )\propto  1/ f$  i.e. slow variations of large amplitude.  Kotov et al. \cite{Kotov et al 2001} improved upon this model and showed that assuming a harder spectrum close to the black hole produces a logarithmic dependence of time-lags on energy, as observed.
The original model by Lyubarskii had no physical description for the viscous fluctuations. A likely mechanism could be magnetic dynamo in the disc. However, dynamo processes vary on a dynamical time-scale  i.e. much faster than the viscous time fluctuations assumed by Lyubarskii.
In a similar framework, King et al. \cite{King et al 2004} suggested that mass accretion rate fluctuations on long time scales could be produced if independent dynamos at a range of disc annuli can 'cooperate'. At each radius a dynamo produces a magnetic field fluctuating  on a dynamical time-scale, both in amplitude and orientation. Very occasionally the magnetic field is aligned over a large surface of the disc. This allows the generation of a local wind or outflow, also causing rapid angular momentum loss in the disc and ultimately a temporary increase of  the mass accretion rate in the inner pat of the disc leading to a X-ray flare occuring on time-scales > $t_{dyn}$. King et al. \cite{King et al 2004}  and then 
Mayer and Pringle \cite{Mayer and Pringle 2006} showed that this kind of models may reproduce observed power spectra. 
The current version of the model assumes that the variability is generated in a geometrically thin disc extending close to the black hole. 
Actually what is observed is the X-ray variability of the corona (either ADC or RIAF).
The model does not explain how the fluctuation of physical parameters in the thin disc affects those in the corona.  Moreover, observations in the soft state have shown that despite the strong coronal activity,  the disc thermal emission is extremely stable. Most of the variability appears to be generated in the corona rather than in the thin disc \cite{Churazov et al  2001}. 

Despite these problems, an interesting feature of this model is that it connects the X-ray timing properties with the formation of outflows, basically predicting a correlation between rapid jet and X-ray activity.
Of course variability of the radio jet on time-scale as short as seconds or less is impossible to observe 
with present radio instrumentation.  However, in some sources, such as XTE J1118+480, the jet has a significant contribution in infrared and optical, making  fast jet photometry possible. 
In  XTE J1118+480, fast optical and UV photometry has revealed a rapid optical/UV
flickering presenting complex correlations with the X-ray variability \cite{Kanbach et al 2001} (see Fig.~\ref{fig:sedxte}). 
This correlated variability cannot be caused by reprocessing 
of the X-rays in the external parts of the disc.
Indeed, the optical flickering occurs on average on shorter
time-scales than the X-ray one \cite{Kanbach et al 2001}, and reprocessing models fail to 
fit  the complicated shape of the X-ray/optical cross correlation 
function.
Spectrally, the jet emission  seems to
 extend at least up to the optical band (see Fig.~\ref{fig:sedxte} and \cite{C03}), 
although the external parts of the disc may provide an
important  contribution to the observed flux at such
wavelengths.
The jet  activity is thus the most likely explanation for the rapid
observed optical flickering. 
For this reason, 
the properties of the optical/X-ray correlation 
in XTE J1118+480 might be of primary importance for the understanding 
of the jet-corona coupling and the ejection process.
One of the most intriguing of these properties is that the two lightcurves seem
 to be related trough a differential relation.
Namely, if the optical variability is representative 
of fluctuations in the jet power output  $P_{\rm j}$, 
the data suggest that the jet power scales roughly like $P_{\rm j} \propto
-\frac{dP_{\rm x}}{dt}$, where $P_{\rm x}$ is the X-ray power \cite{M03}.
Such a relationship is not expected in the model by King et al. \cite{King et al 2004}. On the other hand,  Malzac et al  \cite{Malzac et al 2004} have shown that the
complex X-ray/optical correlations could be understood in terms of an
energy reservoir model.  In this picture, it is assumed that large
amounts of accretion power are stored in the accretion flow before
being channelled either into the jet  or into particle acceleration/ heating in the
Comptonizing region responsible for the X-rays. The optical variability is produced mainly from 
synchrotron emission in the inner part of the jet at distances of a few thousands
 gravitational radii from the hole. It is assumed  that  at any time the optical flux $O_{pt}$ (resp. X-ray flux)
  scales like the jet power $P_{\rm j}$ ( plasma heating power $P_{\rm x}$).

 The essence of the model can be understood using a simple
analogue: Consider a tall water tank with an input pipe and two output
pipes, one of which is much smaller than the other. The larger output
pipe has a tap on it. The flow in the input pipe represents the power
injected in the reservoir $P_{\rm i}$, that in the small output pipe
the X-ray power $P_{\rm x}$ and in the large output pipe the jet power
$P_{\rm j}$.  If the system is left alone the water level rises until
the pressure causes $P_{\rm i}=P_{\rm j}+P_{\rm x}$.  Now consider
what happens when the tap is opened more, causing $P_{\rm j}$ to
rise. The water level and pressure (proportional to $E$) drop causing
$P_{\rm x}$ to reduce. If the tap is then partly closed, the water
level rises, $P_{\rm j}$ decreases and $P_{\rm x}$ increases. The rate
$P_{\rm x}$ depends upon the past history, or integral of $P_{\rm
j}$. Identifying the optical flux as a marker of $P_{\rm j}$ and the
X-ray flux as a marker of $P_{\rm x}$ we obtain the basic behaviour
seen in XTE\,J1118+480.  In the real situation, it is envisaged that the
variations in the tap are stochastically controlled by a shot noise
process. There are also stochastically-controlled taps on the input
and other output pipes as well. The overall behaviour is therefore
complex.

This simple model is largely independent of the physical nature of the energy
reservoir. In a real accretion flow, the reservoir could take the form
of either electromagnetic energy stored in the X-ray emitting region,
or thermal (hot protons) or turbulent motions. The material in the
disc could also constitute a reservoir of gravitational or rotational
energy behaving as described above.
A strong requirement of the model is that the jet power should be at least a few times larger than the X-ray power.  This fact seems to be independently supported by  recent  results on the study of the X-ray radio correlation in black holes and neutron stars indicating that for  black holes in the LHS  the jet power dominates over the X-ray emission \cite{Koerding et al 2006}. Finally this model accounts for the non-linearity: most of the X-ray noise is due to slow fluctuations of the outflow dissipation rate ($P_{\rm j}$ tap) modulating the much faster fluctuations of the coronal dissipation rate ($P_{\rm x}$ tap), and therefore a rms-flux correlation is expected. 

\section{Conclusions}
\label{sec:conc}
In the last decade, a relatively  clear picture of the phenomenology of accretion and ejection in black hole X-ray binaries has emerged. 
However, from a theoretical perspective many of the observed properties are yet to be understood.  The geometry of the accretion flow in the hard state is still debated (ADC versus hot accretion disc). The fundamental physical connection between the hot comptonising plasma and  the compact jet of the LHS remains unclear. Moreover we still have no really satisfactory model for  the triggering of state transitions and the hysteresis of the spectral evolution during outbursts of transients. Finally as it is suspected for XTE~J1118+480, many of the characteristics of the  fast X-ray variability of accreting black holes could be associated to some form of disc jet coupling and remain essentially mysterious. The observed timing properties seem to favours models in which the source of photons is close to the black hole, but the tempo of the variability is dictated by a clock located much farther away.

\end{document}